\documentstyle[preprint,tighten,aps]{revtex}
\def\lapprox{\mathrel{\mathop
  {\hbox{\lower0.5ex\hbox{$\sim$}\kern-0.8em\lower-0.7ex\hbox{$<$}}}}}
\def\gapprox{\mathrel{\mathop
  {\hbox{\lower0.5ex\hbox{$\sim$}\kern-0.8em\lower-0.7ex\hbox{$>$}}}}}

\begin{document}
\author{ B.~Ricci$^{1}$ ,
 	F. L. ~Villante$^{1}$ and
        M.~Lissia$^{2}$,      
         }
\address{
   $^{1}$ Dipartimento di Fisica dell'Universit\`a di Ferrara and
        Istituto Nazionale di Fisica Nucleare, Sezione di Ferrara,
        via Paradiso 12, I-44100 Ferrara, Italy .\\ 
 $^{2}$Istituto Nazionale di Fisica Nucleare, Sezione di Cagliari and 
       Dipartimento di Fisica, Universit\`a di Cagliari,
       Cittadella Universitaria, I-09042 Monserrato, Italy. }

\preprint{\vbox{\noindent
          \null\hfill  INFNFE-01-99}}

\title{Helioseismology and Beryllium neutrinos}

\date{April 1999}

 \maketitle

\begin{abstract}
We derive a lower limit  on the 
Beryllium neutrino flux on earth,
$\Phi(Be)_{min} = 1\cdot 10^9$ cm$^{-2}$ s$^{-1}$,
in the absence of oscillations, 
by using helioseismic data, the B-neutrino flux measured by
Superkamiokande  and the hydrogen abundance at the solar center
predicted by Standard Solar Model (SSM) calculations.
We emphasize that this abundance is the only result of SSMs
needed for getting $\Phi(Be)_{min}$.
We also derive lower bounds for the Gallium
signal, $G_{min}=(91 \pm 3) $ SNU, and for the Chlorine
signal, $C_{min}=(3.24\pm 0.14)$ SNU,
which are about $3\sigma$ above their
corresponding experimental values, $G_{exp}= (72\pm 6)$ SNU
and $C_{exp}= (2.56\pm 0.22) $ SNU.
\end{abstract}

\section {Introduction}
There are several indications, but not really a proof of
solar neutrino oscillations (see e.g. \cite{how,report}): 

\noindent
i) the result of all five solar
neutrino experiments are below the predictions of SSM calculations;
           
\noindent
ii) unless neutrino oscillate, the 
experimental results for Gallium, Chlorine and water detectors
are hardly consistent among themselves, even  if some  experimental result
is discarded;

\noindent
iii) the measured Gallium signal, $G_{exp}= (72\pm 6)$ SNU,
is below the minimal value expected by the luminosity constraint,
$G_{Lum}= (79\pm 2)$ SNU, for standard neutrinos.

Future experiments like Borexino \cite{borexino} and LENS\cite{lens}, 
aiming  at the detection of Be-neutrinos from 
\begin{equation}
\label{ecap}
^7Be+e^- \rightarrow \, ^7Li +\nu_e \quad ,
\end{equation}
can be crucial for the solar neutrino puzzle, since  oscillation schemes
predict  unambiguous seasonal modulations and/or 
drastic reductions with respect to the predicted flux \cite{BP98}:
\begin{equation}
\label{eq_bessm}
\Phi(Be)_{SSM} =4.8 \cdot (1\pm0.09) \cdot 10^9 {\mbox{cm$^{-2}$ s$^{-1}$}} .
\end{equation}
This flux is a very robust prediction of SSM calculation, see \cite{BP98}
for a discussion.  The quoted
$1\sigma$ error is dominated by uncertainty on the $^3He+^4He$ cross section,
as emphasized in  \cite{BP98} and in \cite{TAUP97}. 
This uncertainty alone contributes as much as
all astrophysical uncertainties 
(metal abundance, opacity, luminosity, diffusion...) combined  together.

Nevertheless, eq. (\ref{eq_bessm}) represents the 
outcome of an involved solar model calculation, and it is useful
to provide a lower limit in the
absence of oscillations, $\Phi(Be)_{min}$, 
which
is essentially independent of the SSM. Such is the aim of the present
note.

As well known, Be-neutrinos give a significant contribution to the
Gallium and Chlorine signals, being respectively about one fourth
and one sixth  of the total, according again to the SSM.
The lower limit on $\Phi(Be)$ can thus be used to determine lower limits
to the Gallium and Chlorine signal, in the absence of oscillations.

\section{ A  lower limit to the production rate of Be-neutrinos}

The B-neutrinos from
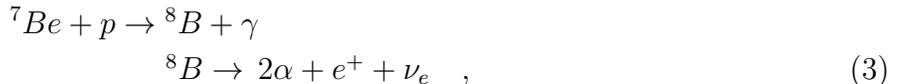
\begin{eqnarray}
\label{pcap}
^7Be+ p  \rightarrow & ^8B + \gamma \quad \qquad \qquad \qquad\nonumber \\
                     & ^8B \rightarrow \, 2\alpha + e^+ + \nu_e \quad ,
\end{eqnarray}
have been observed  by Kamiokande\cite{kam} and Superkamiokande 
\cite{skam}. 
Since B-neutrinos and Be-neutrinos are both sons of 
$^7Be$ nuclei, one expects that detection of the former gives
information on the latter. Our aim is to determine a
lower limit on the production rate of Be-neutrinos, $L(Be)$,
starting from  this consideration.

As the rates of (\ref{ecap}) and (\ref{pcap}) depend differently on the 
solar temperature, we need
some information on it. This is (indirectly) provided by helioseismology,
which determines the sound speed  with
an accuracy of one per cent or better, even close to the solar center, see e.g.
\cite{eliosnoi}. Temperature is obtained from
the sound speed if the chemical composition of the solar plasma is
known. This is the only information which we shall take from SSMs, in the
form of the hydrogen abundance at the solar centre $X_c$, a quantity
which is largely independent of solar models, since it reflects the
amount of hydrogen burnt all along the sun history. Let us make 
this argument in some detail.

The $^7Be$ and $^8B$ luminosities, at production, can be written as:
\begin{eqnarray}
\label{eq1}
L(B)&=& \int d^3r\, n_1 n_7 <\sigma v>_{17} 
    = \lambda_{17}/ T_o^{13}   \int d^3r \, n_1 n_7 T^{13} \\
\label{eq2}
L(Be)&=& \int d^3r\, n_e n_7 <\sigma v>_{e7} 
   = \lambda_{e7}/ T_o^{-0.5}   \int d^3r \, n_e n_7 T^{-0.5}
\end{eqnarray}
where we have used a parametrization 
of the form $<\sigma v>_{ij}=\lambda_{ij}(T/ T_o)^{\alpha_{ij}}$,
the temperature scale $T_o$ is chosen as the central  temperature
of the model in \cite{BP98}, hereafter BP98, $T_o=1.5697\cdot 10^7$ K, 
and  according to \cite{adelberger}: 
\begin{eqnarray}
\label{eql1}
\lambda_{e7}&=& 2.34 \cdot 10^{-33} \, (1 \pm 2\%) 
{\mbox{cm$^3$ s$^{-1}$}} \quad   \\
\label{eql2}
\lambda_{17}&=& 1.04 \cdot 10^{-35} \, (1 \pm 16\%) 
{\mbox{cm$^3$ s$^{-1}$}}\quad   
\end{eqnarray}
(here and in the following errors shown are combinations in quadrature of 
systematic and statistical $1\sigma$ errors).

By using eqs. (\ref{eq1}) and (\ref{eq2}) one can relate the production
rates of Boron and Beryllium neutrinos. From eq. (\ref{eq1}) one has:

\begin{equation}
\label{eqlb}
L(B) = \frac{\lambda_{17}}{ \lambda_{e7}}
 T_o^{-13.5}  \int d^3r n_e n_7 <\sigma v>_{e7} T^{13.5} x
\end{equation}
where $x=n_1/n_e$ is  the ratio of free protons to electrons.
Since temperature decreases  sharply when moving away from the solar center, 
one can assume that at any point in the solar interior
$x(r)T(r)^{13.5}\leq x_c T_c^{13.5}$, where, here and in the
following, the suffix $c$ refers to the solar
center. In this way one has:

\begin{equation}
\label{eqlb2}
L(B) \leq \frac{\lambda_{17}}{ \lambda_{e7}}
\left ( \frac{T_c}{T_o} \right )^{13.5} x_c \, L(Be)  \, .
\end{equation}

As the produced $\nu_e$ can oscillate into species
with a smaller or vanishing cross section in the detector,
the observed luminosity $L(B)_{obs}$ in
Kamiokande and Superkamiokande cannot exceed the produced luminosity,
$L(B)_{obs} \leq L(B)$,
so that one has the following  lower limit  for $L_{Be}$:
\begin{equation}
\label{eqlbe}
L(Be) \geq L(B)_{obs} \frac{\lambda_{e7}}{\lambda_{17}}  
\left ( \frac{T_c}{T_o} \right )^{13.5} 
\frac{1}{x_c} \,.
\end{equation}

Now we use the fact that the solar center
can be described as  perfect gas of fully ionized H and He, to
a very good approximation. In  terms of the isothermal squared
sound speed $u=P/\rho$ and of the hydrogen mass fraction
$X$, this gives:
$     kT_c =  u_c m_p /(3/4 +5/4 X_c)$. One also has:
$     x_c =  2X_c/(X_c + 1) $.
In this way one gets:
\begin{equation}
\label{eqlbe2}
L(Be) \geq
 \frac{\lambda_{e7}}{\lambda_{17}} 
\left (\frac{kT_o}{m_pu_c} \right )^{13.5} 
\frac{(X_c+1)}{(2X_c)}
          (5/4 X_c +3/4)^{13.5}  L(B)_{obs}
\end{equation}

The equation above can of course be translated in terms of fluxes.

We take
 $ \Phi(B)_{obs}=2.42 \cdot (1 \pm 3\% )\cdot 10^6$  cm$^{-2}$ s$^{-1}$
from \cite{skam} and $u_c=1.53 \cdot (1\pm 1\%) \cdot 10^{15} cm^2/s^2$, in 
agreement with helioseismic determinations, see \cite{eliosnoi}.

For the central hydrogen abundance we have to rely on solar model 
calculations.
Recent SSM calculations all yield $X_c$ in the narrow range
$0.333<X_c<0.347$, with a mean value close to the BP98 estimate,
$X_c^{BP98}=0.339$, see table \ref{tabxc}.
The calculated value of $X_c$ is sensitive to opacity, metal
abundance and nuclear  cross sections, see table \ref{tabxc2}. 
The $1\sigma$ 
uncertainty on opacity and metal abundance are respectively
5\% and 6\%, according to \cite{report} and \cite{BP95}, and that on
$S_{pp}$ is 1.7\%, from \cite{adelberger}. By computing
suitable solar models and adding errors in quadrature, we conclude:
\begin{equation}
\label{xcerror}
X_c=0.339 \pm 0.010 \quad .
\end{equation}

We have also computed $X_c$ for a series 
of ``non standard solar models'', where 
some input parameters have been varied, one at the time, by about
$\pm 3 \sigma$ with respect to the SSM reference input, 
see table \ref{tabxc}.
Even in this case, $X_c$ stays in the range 0.329 -- 0.358, i.e. within
about $\pm 5\%$ from BP98.

We remark that $X_c$ is essentially  an indicator
of how much  hydrogen has been burnt so far, starting from an
initial value $X_{in}$ about 0.7. The stability of $ X_c$ corresponds to the fact
that  any solar model has to account for an integrated
solar luminosity of about $L_{\odot}t_{\odot} =5.5 \cdot 10^{50}$ erg.
On these grounds, we consider the adopted value of $X_c$ as rather safe.

In this way we get:
\begin{equation}
\label{eqfibe}
\Phi (Be) \geq (1\pm 0.24) \cdot 10^9 {\mbox{cm$^{-2}$ s$^{-1}$}} \, . 
\end{equation}
where the error include, in quadrature, all uncertainties mentioned
above. The uncertainty on $\lambda_{17}$, $X_c$ and $u_c$ 
contribute to the total error  16\%, 13\% and 12\% respectively.

The inequality (\ref{eqfibe}) defines a minimum flux $\Phi(Be)_{min}
\simeq 1\cdot 10^9$ cm$^{-2}$ s$^{-1}$, which is 
one fifth of the SSM prediction, see eq. (\ref{eq_bessm}).
We note that the only input from SSM 
is the value of $X_c$, whereas all other inputs, 
$\Phi(B)_{obs}$ and $u_c$, are from observational data.

We have obtained this minimal flux using only the physical information that
is relevant to the ratio between the Beryllium and Boron luminosities
$L(Be) / L(B)$; for instance, the actual value of $n_7(r)$ never matters
for our result. However, the additional physical information that determines
the two fluxes separately, in particular the measured $\Phi(B)$, can only
strengthen this limit. In fact, a solar-model-independent analysis of the
Beryllium and Boron neutrino flux production ($0<T_{c}<\infty$, cross sections
more than $3\sigma$'s away from the central values, profiles of densities that
are not constrained by helioseismology varied by factors larger than 30)
shows that the lower limit to $\Phi (Be)$ is
$1.6\times 10^{9}$~cm$^{-2}$~s$^{-1}$, as can be inferred
from Fig.~3 of Ref.~\cite{hemix} and the measured $\Phi(B)$.

\section{ Implications for solar neutrino experiments}

\subsection{Borexino and LENS}

The relevance of  the bound (\ref{eqfibe})
can be appreciated when discussing
the complementarity between Borexino and LENS.
We remind  that LENS is sensitive to $\nu_e$ only, whereas
the signal of Borexino can get contribution also from
$\nu_\mu$ or $\nu_\tau$, their cross section being about 1/5 than 
that of $\nu_e$.

For both experiments, a  signal well below 1/5 of the SSM prediction will be
a definite proof of neutrino oscillations, since it leads to 
a violation of eq. (\ref{eqfibe}).

A signal at the level of 1/5 of the SSM prediction in Borexino
could be interpreted as due to Small Mixing (SM)  angle oscillations 
into active neutrinos, where one expects
that all $\nu_e$ from Beryllium have been transformed into $\nu_\mu$.
However, one could  still insist on standard neutrinos, arguing for
some drastic (maybe desperate) modification of the solar model.
In this situation, a clear discrimination will be provided  by LENS:
for the SM case, the LENS signal,  barrying the background, has to vanish, 
so that the bound (\ref{eqfibe}) will be violated, giving a definite
proof of neutrino oscillations (furthermore, the comparison with
Borexino will show the presence of $\nu_{\mu}$ or $\nu_{\tau}$).

\subsection{Gallium experiments}

As well known, the solar luminosity essentially fixes the total
production rate of neutrinos.
 Since neutrino cross sections increase
with energy, the minimal Gallium signal, in the absence of oscillations,
can be estimated by assuming that the total flux consists of pp neutrinos 
only. The pep-neutrinos can be safely included in this estimate, as
the ratio of pep to pp-neutrinos is well known and essentially
unsensitive to solar physics details, see \cite{primo} and
\cite{report}.
By using  updated cross sections from \cite{cross}, this
arguments gives as a minimal Gallium signal in the absence of oscillations:
\begin{equation}
\label{eqsgalum}
           G_{Lum} = (79.5 \pm 2.0) \, {\mbox{SNU}}
\end{equation}
where the error arises mainly
from the capture cross section of pp neutrinos.

As well known, also B-neutrinos contribute to the Gallium signal.
Their contribution is best estimated by using experimental data.
If one takes  into account the flux measured by Superkamiokande, 
with the capture  cross section of \cite{cross},
this contributes  an additional $(5.8 \pm 1.5)$ SNU,
where most of the error comes again from the capture  cross section.
All this results in:
\begin{equation}
\label{sgalum2}
         G_{Lum+SK}= (85.3 \pm 2.5) \, {\mbox{SNU}} \, .
\end{equation}

According to the previous discussion, one has to include
now the minimal contribution of $^7Be$ neutrinos. 
For $\Phi(Be)=\Phi(Be)_{min}$ 
by using the luminosity constraint (see section 2.4 of \cite{report}) 
one has an additional contribution
of $(5.9 \pm 1.4)$ SNU, where most of the error comes from $\lambda_{17}$, 
so that in conclusion the minimal Gallium signal is now:
\begin{equation}
\label{eqsgamin}
         G_{min}= (91 \pm 3) \,{\mbox{SNU}} \,.
\end{equation}
This has to  be compared with the Gallex \cite{gallex} and Sage\cite{sage}
 average:
\begin{equation}
\label{eqsgaexp}
         G_{exp}= (72 \pm 6) \, {\mbox{SNU}} \,.
\end{equation}

All this means that the present  experimental result is about three sigmas
below the minimal expectation in the absence of oscillations.

This also illustrates the potential of GNO \cite{gno}, the 
successor of Gallex, which should reduce the total error down to 
about 4 SNU. If the present central value is mantained, the
discrepancy with the minimal prediction will be at the level of about
$5\sigma$, thus providing a clean signature of neutrino oscillations.

\subsection{The Chlorine result}

The solar luminosity constraint, together with
the assumption that the ratio of pep-neutrinos over
pp-neutrinos $\psi=\Phi(pep)/ \Phi(pp)=0.0023$
is correctly determined by SSM calculation, can be used to provide a lower 
limit also for the Chlorine signal.

As well known, due to the fact that cross sections increase
with neutrino energy, the minimal signal is obtained by maximixing the number
of lowest-energy neutrinos, consistent with the luminosity constraint. 
Since a flux  $\Phi_{pp}$ of pp-neutrinos is anyhow accompanied
by a flux $\Phi(pep)= \psi \Phi(pp)$, this implies a minimal 
Chlorine signal:
\begin{equation}
\label{cl1}
C_{Lum}=\frac{K_\odot}{Q_{pp}/\psi +Q_{pep}} \,  \sigma_{pep} \, ,
\end{equation}
where $K_{\odot}$ is the solar constant, $Q_i$ is the average electromagnetic
energy released for emitted $i$-neutrino (see \cite{report}) and 
$\sigma_{i}$ is the averaged $i$-neutrino cross section on Chlorine
detector. By using the cross section from \cite{libro},
but for the
absorption cross section of $^{8}B$ neutrinos  from 
\cite{sboro}, one has:
\begin{equation}
\label{cl2}
C_{Lum}= (0.243 \pm 0.005 )\, {\mbox{SNU}} \, .
\end{equation}

From the Superkamiokande result one can deduce the B-neutrinos contribution
of $(2.76 \pm 0.12)$ SNU, so that:
\begin{equation}
\label{cl3}
C_{Lum+SK}= (3.0 \pm 0.1 )\, {\mbox{SNU}} \, .
\end{equation}
The minimal Be-neutrino flux  implies  the additional
contribution of
\begin{eqnarray}
\label{cl4}
C_{Be_{min}}&=& \Phi(Be)_{min} [ \sigma_{Be} - \sigma_{pep}
\frac{Q_{Be}}{Q_{pp}/\psi +Q_{pep}}] \nonumber \\
            &=& (0.24 \pm 0.06) \, {\mbox{SNU}} \ .
\end{eqnarray}
In this way the minimal Chlorine signal becomes:
\begin{equation}
C_{min}=(3.24 \pm 0.14)  \, {\mbox{SNU,}}
\end{equation}
to be compared with the experimental result \cite{homestake}
\begin{equation}
C_{exp}=(2.56 \pm 0.22)  \, {\mbox{SNU.}}
\end{equation}

Again the Be-neutrinos contributions, 
eq. (\ref{cl4}), corresponds to the ``$1\sigma$``
uncertainty of the experimental result, and again the experimental
signal is about three sigmas below the minimal prediction.

\subsection{Combining experimental results}

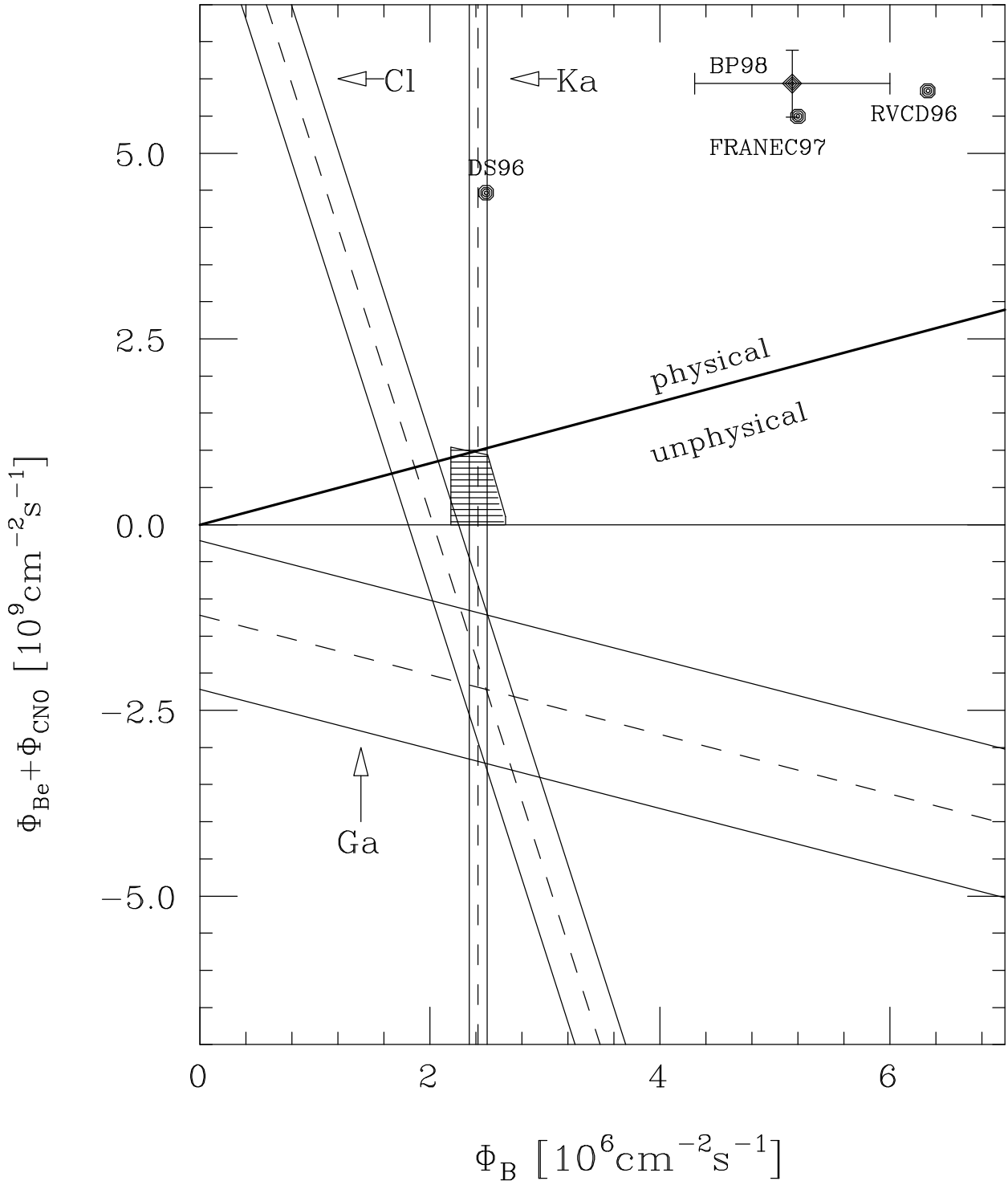
\begin{figure}
\caption[aff]{
The $^8B$ and $^7Be+CNO$ neutrino fluxes, 
consistent with the luminosity constraint
and experimental results for standard neutrinos.
The dashed (solid) lines correpond to the central ($\pm 1\sigma$) experimental
values for Chlorine, Gallium 
and $\nu - e$ scattering  experiments. 
The dashed area corresponds to the region within $3\sigma$ from each
experimental result.
The predictions of solar models including element diffusion (full circles)
~\cite{DS96,BP95,RCVD96,ciacio},
are also shown.
The thick diagonal line corresponds to the limit on $\Phi(Be)$ derived
in this paper, see eq. (\ref{eqbebo}).
}
\label{fig1}
\end{figure}

A global view of the ``solar neutrino puzzle`` is presented in Fig. 
\ref{fig1} which updates Fig. 7 of \cite{report}.
As a generalization of eq. (\ref{eqfibe}) for an
arbitrary value of the observed $^8B$ flux 
$\Phi(B)$ one has:
\begin{equation}
\label{eqbebo}
\Phi(Be)_{min}=4\cdot 10^2 \Phi(B) \, .
\end{equation}
The corresponding 
thick ``diagonal`` line,  in Fig. 1
 defines thus the lower border
of the physical region. The shaded area, corresponding to the region 
within $3 \sigma$ from each experimental result, is almost
completely out 
of  the physical region.

\section{Conclusions}

As a summary, in the absence of oscillations we predict:
\begin{eqnarray}
\Phi(Be)_{min} &=& (1\pm0.24)\cdot 10^9\, 
{\mbox{cm$^{-2}$ s$^{-1}$}} \nonumber \\
 G_{min}&=&(91 \pm 3 )\, {\mbox{SNU}} \nonumber \\
 C_{min}&=&( 3.24\pm 0.14 )\, {\mbox{SNU}}  \nonumber \, .
\end{eqnarray}
Let us list the information and assumptions behind 
these results:

\noindent
i)  the measured $^8B$ flux by Superkamiokande;

\noindent
ii) the helioseismically determined sound speed, near or
    at the solar center, $u_c$;

\noindent
iii) the measured  value of $\lambda_{17}$;

\noindent
iv) the value of $\lambda_{e7}$, derived from
the lifetime of $^7Be$ in the laboratory;

\noindent
v)  the luminosity constraint, i.e. the present observed solar luminosity
    equals the presently generated nuclear power in the sun;

\noindent
v)  the central Hydrogen  abundance, the only information  we take
    from SSM calculations.

We remark that  we do not need to know the central solar temperature,
nor the values of the astrophysical S-factors for the He+He reactions.

\acknowledgments

We are extremely grateful to V. Berezinsky and G. Fiorentini for suggesting
us the problem and for useful discussions and comments.

\begin{table}
\caption[a]{Values of $X_c$ in standard and non-standard solar models}
\begin{tabular}{lc}
\multicolumn{2}{c}{standard solar models}\\
BP95 \cite{BP95} & 0.3333\\
BP98 \cite{BP98} & 0.3387 \\
RCVD96 \cite{RCVD96} & 0.3328 \\
DS96 \cite{DS96} & 0.3424 \\
TC98\cite{TC98} & 0.3442 \\
FR97 \cite{ciacio} & 0.3467 \\
\hline
\multicolumn{2}{c}{non standard solar models}\\
$S_{pp}\times0.9$ & 0.3414 \\
$S_{pp}\times1.1$ & 0.3354 \\
$opacity \times  0.9$    & 0.3647 \\
$opacity \times1.1$    & 0.3287 \\
$Z/X \times 0.9$    & 0.3579 \\
$Z/X \times 1.1$    & 0.3354
\end{tabular}
\label{tabxc}
\end{table}

\begin{table}
\caption[b]{Dependence of $X_c$ on the solar model inputs $P_i$}
\begin{tabular}{lc}
$P_i$ & $ \partial ln X_c/ \partial P_i$ \\
\hline
$S_{pp}$ & 0.1 \\
$opacity$  & -0.5 \\
$Z/X$ & -0.3
\end{tabular}
\label{tabxc2}
\end{table}

\end{document}